\documentstyle[epsf]{elsart}

\begin{document}
\begin{frontmatter}

\title{Two Simple Approaches to Sol-Gel Transition}

\author{Alexander I. Olemskoi\thanksref{AOVB}}
\address{
Sumy State University\\
Rimskii-Korsakov St.  2, Sumy, 40007 Ukraine \\
}
\author{Ivan Krakovsk\'{y}}
\address{
Department of
Macromolecular Physics, Faculty of Mathematics \& Physics\\
Charles University\\
V Hole\v{s}ovi\v{c}k\'{a}ch 2, 180~00 Prague 8, Czech Republic\\
}

\thanks[AOVB]{olemskoi\char'100ssu.sumy.ua, alexander\char'100olem.sumy.ua}

\begin{abstract}

We represent a theory of polymer gelation as an analogue
of liquid-glass transition in which elastic fields of stress and strain shear
components appear spontaneously as a consequence of the cross-linking
of macromolecules.  This circumstance is explained on the basis of
obvious combinatoric arguments as well as a synergetic Lorenz
system, where the strain acts as an order parameter, a conjugate
field is reduced to the elastic stress, and the number of cross-links
is a control parameter.  Both the combinatoric and synergetic
approaches show that an anomalous slow dependence of the shear
modulus on the number of cross-links is obtained.

\vspace{0.5cm}


\vspace{0.5cm}


\end{abstract}

\end{frontmatter}

\section{INTRODUCTION}

Within the phenomenological framework the basic
distinction between a liquid and a glass consists in character of the
relaxation law of shear components of the elastic stress: if in
an ideal glass they are kept infinitely long, in a liquid such
relaxation proceeds in the finite time $\tau =
\eta /G$, where $\eta$ is the dynamical shear
viscosity and $G$ is the shear modulus \cite{1}. In naive manner it
is possible to assume, that a glass transition is caused purely by
kinetic effect of a liquid freezing, for which viscosity $\eta$ gets
infinite value for a finite value of the shear modulus $G$ \cite{2}.
However in the course of usual second-order phase transition, where
infinite increase of the $\tau$ at critical point is also observed
the situation is reverse.  Really, proceeding from the viscoelastic
liquid to a general case, one has $\tau = \chi / \gamma$, where
$\chi$ is the generalized susceptibility and $\gamma$ is the kinetic
coefficient (in the case under consideration they are reduced to the
quantities $G^{-1},~\eta^ {-1}$, respectively) \cite{3}.  The
infinite increase of a susceptibility $\chi$ occurs and a kinetic
coefficient $\gamma$ does not manifest any peculiarity at the
phase transition.  In our case, this is equivalent to the situation
when the shear modulus $G$ is approaching zero value at a fixed
viscosity $\eta$.  This situation corresponds to a viscoelastic
transition \cite{4}.

Usually under a glass transition, the transition into a glassy state which
occurs during fast cooling of a liquid is understood. However, an analogous
type of transition occurs during polymer network formation ({\em polymer
gelation}) from monomers or linear macromolecules ({\em sol}) by means of a
chemical reaction at constant temperature. This transition is referred to as
the {\em sol-gel transition}. In such a case, thermodynamic peculiarities are
observed as, e.g., appearance of the shear modulus when the system reaches a
critical point.\footnote{ Of course, when the temperature of the polymer
network formed is decreased the kinetic glass transition in above sense occurs,
as well.} The first theoretical representations of such type transitions have
been elaborated in classical works by Flory \cite{flory1} -- \cite{flory3}.
Modern description of the glassy state of the macromolecule
networks (see \cite{G} -- \cite{E}) is based on pioneering contributions
to the theory of soft condensed matter:
the Deam-Edwards theory \cite{DE} of a cross-linked tangled macromolecule,
and the Edwards-Anderson theory \cite{EA} of spin glass.

The aim of this paper is to elaborate the simplest theoretical scheme
so that the transition occurring during polymer gelation
can be explained on the base of obvious combinatoric arguments
(Section 2) as well as within framework of a synergetic theory
(Section 3). In the former case, the critical point and behaviour of
the system behind the critical point are obtained using information
about the cross-linking properties of the constituent parts of the
system ({\em monomers}) and extent of the cross-linking process,
only.  In Section 3, the starting point is that a sol-gel transition
is ensured by self-organization of the elastic fields of stress
and strain shear components and the number of
cross-links.  Both the combinatoric approach and the synergetic
one allow to obtain anomalous dependence of the shear modulus on the
number of cross-links.

\section{Combinatoric approach}

The gelation process has been described first
by Flory\cite{flory1} and Stockmayer\cite{stock}.
They used a laborious approach based on combinatoric considerations
of the most probable composition of the system.
A much more effective variant of this approach,
exploiting theory of branching processes,
has been developed later by Dobson and Gordon \cite{dg1,dg2}.
This approach allows also an analysis of structural details
of the system behind the critical point, see, e.g. \cite{dusek}.

As the simplest case let us consider the gelation process
in the system
which at the beginning consists of a large number of monomers
$\cal N$, wearing $f$ functional terminal groups
of independent reactivity.
Monomers react mutually and irreversibly
via their terminal groups and if $f>2$,
branched molecules of increasing size and complexity
are formed progressively in the system.
Extent of the reaction is described
by the conversion of terminal groups $\alpha$,
which is defined as the ratio of the number of groups
consumed by the reaction at given time and the starting number of groups.
Eventually, when a critical conversion $\alpha_{\rm c}$
is achieved a molecule of macroscopic dimensions ({\em gel}) appears
in the system {\cite{flory1}, \cite{flory2}}.
Gel contains cycles of long sequences of linked monomers
and, consequently, attains elastic properties.
Flory \cite{flory3} has shown that shear modulus of gel $G$
is proportional to its ``cycle rank'' $\xi$
defined as the number of ``superfluous'' links formed in gel
which can be cut without breaking integrity of gel.
In other words, cycle rank is the number of cuts needed
for elimination of all cycles from gel.

The key role in the approach is played by the extinction probability $v$,
which is a probability that a link formed in the process has
just a finite continuation.
It can be shown easily \cite{dg1} that for the system considered here
the extinction probability can be obtained as a root of the equation
\begin{equation}
v=(1-\alpha+\alpha v)^{f-1}
\label{v1}
\end{equation}
satisfying the condition $0\le v\le 1$.
Eq.(\ref{v1}) expresses fact that a given link has a finite
continuation in a direction only if $f-1$ functional groups remaining
on the monomer connected by the link are either unreacted
(with probability $1-\alpha$) or reacted giving rise to links with
finite continuation only (with probability $\alpha v$).
Below the critical conversion, the only solution of Eq.(\ref{v1})
is $v=1$, i.e., only molecules of finite size
are formed in the system.
However, behind the critical conversion
monomers can be found either in sol or in gel: $v<1$.
As a measure of the ``distance'' from the critical conversion
let us introduce the parameter $\epsilon$ defined as
\begin{equation}
\epsilon\equiv\alpha-\alpha_{\rm c}
\label{eps}
\end{equation}
and expand the extinction probability in a series:
\begin{equation}
v=1+A_{1}\epsilon+A_{2}\epsilon^{2}+\ldots
\label{v3}
\end{equation}
Substituting Eqs.(\ref{eps}) and (\ref{v3}) into Eq.(\ref{v1}),
expressions for the critical conversion and the parameters $A_{i}$
are obtained as
\begin{eqnarray}
\alpha_{\rm c}&=&\frac{1}{f-1}\label{alpha},\\
A_{1}&=&-2~\frac{(f-1)^{2}}{f-2}\label{a1},\\
A_{2}&=&\frac{4}{3}~\frac{(2f-3)(f-1)^{3}}{(f-2)^{2}}.
\label{a2}
\end{eqnarray}

The expression for the cycle rank can be found in the following
way.  Obviously, to join $\cal N$ monomers into a cycle-free
structure, ${\cal N}-1$ links are necessary. By definition, the cycle
rank $\xi$ is the number of ``superfluous'' links formed in gel,
i.e., the difference between the number ${\rm N}_{\rm G}$ of all
links formed in gel and the number ${\cal N}_{\rm G}-1$ of links
sufficient to join together ${\cal N}_{\rm G}$ monomers of gel:
\begin{equation} \xi\equiv {\rm N}_{\rm G}-({\cal N}_{\rm G}-1)\simeq
{\rm N}_{\rm G}-{\cal N}_{\rm G} \label{xi1} \end{equation} as ${\rm
N}_{\rm G}$, ${\cal N}\gg 1$.  Correspondingly, the number of links
formed in gel is the difference between the numbers of links formed
in the total system and in sol, i.e., \begin{equation} {\rm N}_{\rm
G}=\frac{1}{2}{\cal N}\alpha f(1-v^2) \label{ng} \end{equation} as any
link in sol has to have finite continuations in two directions.  On
the other hand, the number ${\cal N}_{\rm G}$ of monomers
incorporated in gel is the difference between the total number of
monomers and the number of monomers in sol which is made of monomers
with links of only finite continuation:  \begin{equation} {\cal
N}_{\rm G}={\cal N}-{\cal N}(1-\alpha+\alpha v)^{f}.  \label{Ng}
\end{equation} By virtue of Eqs.(\ref{xi1}) -- (\ref{Ng}),
one gets for cycle rank of the system considered
\begin{equation} \frac{\xi}{\cal N} = (1-\alpha+\alpha
v)^{f}+\frac{1}{2}\alpha f(1-v^{2})-1.  \label{xi2} \end{equation}
Finally, substituting Eqs.(\ref{v3}),(\ref{a1}) and (\ref{a2})
into the formula (\ref{xi2}), one gets the necessary expansion:
\begin{equation}
\frac{\xi}{\cal N} =
\frac{2}{3}~\frac{(f-1)^{4}}{(f-2)^2}f~\epsilon^3+O(\epsilon^{4}).
\label{xi3}
\end{equation}
Respectively, the weight fraction of gel
$w_{\rm G}\equiv{\cal N}_{\rm G}/{\cal N}$
is determined by Eq.(\ref{Ng}) to read
\begin{equation}
w_{\rm G}=2~\frac{(f-1)^{2}}{f-2}~\epsilon+O(\epsilon^{2}).
\label{wg2}
\end{equation}
So, if the critical exponent of the gel weight fraction is equal 1
as usual, this for the cycle rank is anomalous large being equal 3.
It is worthwhile to note that combinatoric approach is based
exclusively on the information about the functionality of the
monomers and extent of the chemical reaction between the monomers in
the system considered.

\section{Synergetic approach}

Now, let us consider the polymer network as a viscoelastic continuum
matter that is characterized by the shear modulus $G$ and the
shear viscosity $\eta$. Process of polymer gelation is determined
by the number of the cross-links $N$, which value is different from a
stationary magnitude $N_0$ at a time $t$.
Therein, an elastic state of the polymer is defined
by the shear component  of the proper (internal) values
of deformation $\varepsilon(t)$ and stress $\sigma(t)$.
The keypoint is that these values are not reduced to the external
elastic deformation $e\ll 1$ and stress $\sigma_e\ll G$,
in particular they can get large magnitudes $\varepsilon\sim 1$,
$\sigma\sim G$.

Our consideration of evolution of the elastic continuum with the
internal structure is stated on the phenomenological equations by
Maxwell-Kelvin \cite{1}
\begin{equation}
{d\varepsilon\over dt}=-{\varepsilon\over \tau}+{\sigma\over\eta},
\label{a}
\end{equation}
\begin{equation}
{d\sigma\over dt}=-{\sigma\over \tau_{\sigma}}+
g_{\sigma}\varepsilon N.
\label{b}
\end{equation}
Here we introduce a macroscopic relaxation time $\tau$ for
the strain and a microscopic one $\tau_{\sigma}$ for the stress,
as well as a constant $g_{\sigma} > 0$ of the positive feedback between
the deformation $\varepsilon$ and the number of cross-links $N$.
Within the microscopic interval $t\gg\tau_{\sigma}$,
steady-state condition $d\sigma/dt=0$ in Eq.(\ref{b}) leads to
the Hooke law with the microscopic shear modulus
\begin{equation}
G_{\sigma}\equiv\tau_{\sigma} g_{\sigma}N
\label{c}
\end{equation}
being determined by the number of cross-links $N$.
Respectively, within a macroscopic interval $t\gg\tau$
Eq.(\ref{a}) gives the magnitude
$G\equiv\eta/\tau$
that is characteristic for the usual modulus of the
viscoelastic matter.  Lastly, a variation rate $dN/dt$ of the
internal degree of freedom is supposed to be determined by the
equation \begin{equation} {dN\over dt}={N_0-N\over
\tau_N}-g_N\sigma\varepsilon \label{e} \end{equation} where $\tau_N$
is a mesoscopic relaxation time, $g_N > 0$ is constant of negative
feedback between the deformation $\varepsilon$ and the stress
$\sigma$.  Within a mesoscopic interval $\tau_N\ll t\ll\tau$,
Eq.(\ref{e}) determines a steady-state value
\begin{equation}
N=N_0-\tau_N g_N\sigma\varepsilon
\label{f}
\end{equation}
that is smaller than the magnitude $N_0$ fixed by external conditions
due to the fact that the elastic energy is proportional to
the product $\sigma\varepsilon$.

System of Eqs.(\ref{a}), (\ref{b}) and (\ref{e}) is known in
synergetics \cite{Haken} as the Lorenz system where the deformation
$\varepsilon$, the stress $\sigma$ and the number of cross-links $N$
play roles of an order parameter, a conjugate field and a control
parameter, respectively.
It is very important for following considerations that the
relation between micro-, meso- and macroscopic values of
the relaxation times
\begin{equation} \tau_{\sigma}, \tau_N \ll
\tau \label{g} \end{equation}
is satisfied. Due to this condition
the evolution of the quantities $\sigma$, $N$ turns out to
be subordinated to the long-time variation of $\varepsilon$.
A peculiarity of the Lorenz system consists in linear character
of the equation (\ref{a}) for the order parameter
$\varepsilon$ and in non-linearity of equations (\ref{b}),
(\ref{e}) for the conjugate field $\sigma$ and the control parameter
$N$.  The negative nature of non-linearity in Eq.(\ref{e}) means a
decrease of the number $N$ of cross-links.  Evidently, this fact
reflects Le Chatelier principle.  A non-linear term in Eq.(\ref{b})
for a field $\sigma$ describes the positive feedback causing the
system self-organization.

Expressions (\ref{a}), (\ref{b}) and (\ref{e}) form
the complete system of equations
determining the polymer cross-linking behaviour.
Because of a slow evolution, the order
parameter $\varepsilon(t)$ subordinates
variations of quantities $\sigma (t)$, $N(t)$,
so that one can take $d\sigma/dt =dN/dt =0$
within the framework of the adiabatic approximation \cite{Haken}.
Then $N$, $\sigma$ are expressed
in terms of $\varepsilon$ by the equations:
\begin{equation}
N={N_0\over 1+
\varepsilon^2/
\varepsilon^2_m},\quad
\varepsilon_m^{-2}\equiv \tau_{\sigma} \tau_N g_{\sigma} g_N;
\label{h}
\end{equation}
\begin{equation}
\sigma=G_0{\varepsilon \over 1+
\varepsilon^2/
\varepsilon^2_m},\quad
G_0 \equiv \tau_{\sigma} g_{\sigma} N_0.
\label{i}
\end{equation}
In accordance with
Eq.(\ref{h}) the number of cross-links $N$ decreases
monotoneously with increase of the strain $\varepsilon$ from
the value $N_0$ at $\varepsilon = 0$ to $N_0/2$ at $\varepsilon
=\varepsilon_m$.\footnote{Obviously this decrease is caused by the
negative feedback in Eq.(\ref{e}), that is reflection of Le Chatelier
principle for analyzes problem. Actually,  a liquid
self-organization, resulting in a sol-gel transition, is caused by the
positive feedback between the strain $\varepsilon$ and the number of
cross-links $N$ in Eq.(\ref{b}).  Hence, the increase of the number
of cross-links should intensify the self-organization effect.
However, according to Eq.(\ref{h}) system behaves in such way that
the consequence of self-organization, i.e., growth of the elastic
strain, leads to decrease of its origin -- the number of
cross-links.} In Eq.(\ref{i}) the elastic stress in terms of the
strain has the linear form of the Hooke law at $\varepsilon\ll
\varepsilon_m$ with the effective shear modulus $G_0$. Then, at
$\varepsilon = \varepsilon_m$ the function $\sigma (\varepsilon) $
has a maximum and at $\varepsilon > \varepsilon_m$ it decreases which
is physically meaningless. Thus, the constant $\varepsilon_m$,
defined by the second equation~(\ref{h}), has the meaning of the
maximum achievable strain.

Substituting Eq.(\ref{i}) in Eq.(\ref{a}) we find
the equation describing evolution of a system in the course of the
sol-gel transition:  \begin{equation} {d\varepsilon\over dt}=-\gamma
{\partial E \over \partial\varepsilon},\qquad
\gamma\equiv {\varepsilon_m^2\over\tau T~N_0}
\label{j}
\end{equation}
where constant $\gamma$ plays a role of the kinetic coefficient.
Behaviour of the system under consideration is determined by
the dependence $E (\varepsilon) $ of the elastic energy on the strain:
\begin{equation} E \equiv {T~N_0\over 2}
\left [{\varepsilon^2 \over \varepsilon^2_m} - {N_0\over N_c}
\ln\left(1+{\varepsilon^2\over
\varepsilon^2_m}\right)\right]
\label{k}
\end{equation}
where the characteristic value of the number of cross-links is introduced
\begin{equation}
N_c \equiv {\eta\over\tau_{\sigma} g_{\sigma}}.
\label{l}
\end{equation}
At $N_0\le N_c$ dependence (\ref{k}) is monotoneously
increasing with a minimum at the point $\varepsilon = 0$. It
means that in the stationary state ($\dot\varepsilon {=} 0$) the
elastic strain is not spontaneous. Thus, a liquid state is
realized, in which the strain caused by the external stress relaxes
during the time
\begin{equation} \tau_ {ef} = \tau \left (1 -
N_0/N_c\right)^{-1}.
\label{m}
\end{equation}
The relaxation time
increases infinitely when the number of cross-links $N_0$
reaches the critical value $N_c$ and at $N_0> N_c$ the system
undergoes a sol-gel transition. In the gel state  the multiplier 1/2
appears in the dependence (\ref{m}), and the minimum of the
elastic energy corresponds to the elastic strain
\begin{equation}
{\varepsilon^2 \over \varepsilon^2_m} = {N_c
\over N_0}~{N_0-N_c\over N_c}\equiv {\epsilon \over 1 +
\epsilon}
\label{n}
\end{equation}
where we introduce the distance
from the critical value $N_c$ \begin{equation} \epsilon\equiv
{N_0-N_c\over N_c} \label{o} \end{equation}
being equivalent to the
definition (\ref{eps}).  Inserting Eq.(\ref{m}) into the dependence
(\ref{k}), we obtain the elastic energy of the steady-state:
\begin{equation}
E_0\equiv E(\varepsilon_0)=-~{TN_0\over
4}~\epsilon^2 + O(\epsilon^3).
\label{p}
\end{equation}
As would be expected, this
energy is proportional to the second power of the parameter
(\ref{o}) and is negative in nature (the latter
means the energy benefit of the gel state in comparison
with the liquid state).

Taking into account that the glassy state is determined by density
of the localized monomers, let us find now the shear modulus of the appeared
gel state.  It is principally important in our
considerations that the gel state is determined by the value of
the Deam-Edwards parameter of localization $\omega$ \cite{DE} which is
supposed to be proportional to the square of the proper strain
$\varepsilon_0$ of the matter.  Under the condition of the
appearance of the elastic strains $e$,
a generalized Deam-Edwards parameter $\omega(e)$ has to be
considered which is related to the  condition $\omega(e=0)\equiv \omega$.
Then, expanding the function $\omega(e)$ into a series and keeping the first
two terms only, one obtains
\begin{equation} \omega(e)\simeq
\omega(1+e^2), \quad \omega\propto\varepsilon^2_0,\quad e\ll 1.
\label{r1} \end{equation}
By virtue of the parity condition $\omega(e)=\omega(-e)$, 
this expansion does not contain a linear term.\footnote{
Otherwise, the usual squared dependence (\ref{u}) will not be obtained.}
Because of that the total strain
$\varepsilon\propto\sqrt{\omega(e)}$, internal one
$\varepsilon_0\propto\sqrt{\omega}$ and elastic one $e$ are connected
by the following relation:  \begin{equation} \varepsilon^2 =
\varepsilon_0^2 (1+e^2), \qquad e\ll 1.  \label{r} \end{equation}
The key point is that Eq.(\ref{r}) supposes the additivity rule
holds not for quantities $\varepsilon$, $\varepsilon_0$, $e$
themselves, but for their squares.  The physical reason for such a
situation is that the system under consideration is random in
character and described by a symmetrical distribution function.
Therefore, all odd-power moments vanish identically and making use
of the additivity rule (\ref{r}) for variances follows.

It is worthwhile to note the seeming contradiction between above relations
for the localization parameter $\omega$
and the Deam-Edwards results \cite{DE}. Obvious reason consists in that
the formers are obtained within framework
of the mean-field theory, whereas the latters suppose 
fluctuation effects.
According to \cite{DE} the cross-linking process
is not sensible to strain $\varepsilon$ 
and corresponding localization parameter $\omega$
is proportional to the cross-link number $N$ but not to the
difference $N-N_c$, as in Eqs.(\ref{n}), (\ref{o}). In our opinion, this
is caused by non-self-consistency of the approach \cite{DE} in sense
that the stress field $\sigma$ is switched off.  On the other hand,
making use of the statistical scheme \cite{DE} arrives at the
strain-dependence for the  localization parameter $\omega$ due to
appearance of the polymer network entanglement, whereas
within the above phenomenological approach this dependence 
has to be postulated.

A contribution to the elastic energy caused by the
external strain is determined by equality \begin{equation} \Delta
E(e)\equiv | E\left(\varepsilon(e)\right)- E(\varepsilon_0)|.
\label{s}
\end{equation}
With use of Eqs. (\ref{k}), (\ref{p}), (\ref{r}) and expansion
$\ln(1+x)\approx x-x^2/2+x^3/3$,~$ x\ll 1$
the expression
\begin{equation}
\Delta E(e)\approx N_0~{T\epsilon^3\over 2} e^2
\label{u}
\end{equation}
is easily obtained where only the first non-zero term is
kept.  Comparing this relation with the usual expression for the
elastic energy \cite{1} \begin{equation} \Delta E(e)\equiv V~{G\over
2} e^2 \label{v} \end{equation} where $V$ is volume, we arrive at
the final expression for the shear modulus of the gel
state of the polymer network:
\begin{equation}
G={T\epsilon^3\over\Omega},
\quad \epsilon\equiv {N_0-N_c\over N_c}\ll 1,
\quad \Omega\equiv
{V\over N_0}.
\label{w}
\end{equation}
The notable peculiarity of
this result consists in that, in accordance with previous result
(\ref{xi3}), the shear modulus is proportional to the third power of
the distance $\epsilon$.  It is worthwhile to note that an expression
of this kind can be obtained only within framework of the synergetic
approach used, but not on the basis of the phase transition theory by
Landau.

Finally, comparing Eqs. (\ref{xi3}) and (\ref{w}), we obtain the
relation between the micro- and macroscopic parameters of the gel
under consideration in the proximity
of the critical point ($\epsilon\ll 1$):
\begin{equation}
G=g~\frac{\xi}{\cal N}~\frac{T}{\Omega},\quad
g\equiv\frac{3}{2}~f^{-1}~\frac{(f-2)^2}{(f-1)^{4}}.
\label{xy}
\end{equation}


\begin{thebibliography} {00}

\bibitem{1} L.~D.~Landau, E.~M.~Lifshitz, {\it Elasticity Theory}
(Nauka, Moscow, 1987).

\bibitem{2} A.~Havranek, M.~Marvan, Ferroelectrics {\bf 176}, 25
(1996).

\bibitem{3} A.~I.~Olemskoi, {\it Theory of Structure Transformations in
Non-equilibrium Condensed Matter} (NOVA Science, N.-Y., 1999).

\bibitem{4} V.~G.~Bar'yakhtar, A.~I.~Olemskoi, Sov. Phys. Solid State
{\bf 33}, 2705 (1991).

\bibitem{flory1} P.~J.~Flory, J.Am.Chem.Soc. {\bf 63}, 3083 (1941).

\bibitem{flory2} P.~J.~Flory, {\it Principles of Polymer Chemistry}
(Cornell University Press, 1953).

\bibitem{flory3} P.~J.~Flory, Proc.Roy.Soc.London A{\bf 351}, 351
(1976).

\bibitem{G} P.~M.~Goldbart, H.~E.~Castillo, A.~Zippelius,
Adv. Phys. {\bf 45}, 393 (1996).

\bibitem{P} S.~Panyukov, Y.~Rabin, Phys. Rep. {\bf 269}, 1 (1996).

\bibitem{E} S.~F.~Edwards, H.~Takano, E.~M.~Terentjev, cond-mat/0007270.

\bibitem{DE} R.~T.~Deam, S.~F.~Edwards, Phil.Trans.R.Soc. A{\bf 280},
317 (1976).

\bibitem{EA} S.~F.~Edwards, P.~W.~Anderson, J.Phys. F{\bf 5}, 965
(1975).

\bibitem{stock} W.~H.~Stockmayer, J.Chem.Phys. {\bf 11}, 45 (1943).

\bibitem{dg1} G.~R.~Dobson, M.~Gordon, J.Chem.Phys. {\bf 41}, 2389
(1964).

\bibitem{dg2} G.~R.~Dobson, M.~Gordon, J.Chem.Phys. {\bf 42}, 705
(1965).

\bibitem{dusek} K.~Du\v{s}ek, Adv.Polym.Sci. {\bf 78}, 1 (1986).

\bibitem{Haken} G.~Haken, {\it Synergetics} (Springer-Verlag,
Berlin, Heidelberg, N.Y., 1983).

\end{thebibliography}
\end{document}